\documentstyle[12pt,graphicx,subfigure]{article}
\def\slash#1{\setbox0=\hbox{$#1$}#1\hskip-\wd0\hbox to\wd0{\hss\sl/\/\hss}}
\begin{document}
\baselineskip=20 pt
\def\l{\lambda}
\def\m{\mu}
\def\L{\Lambda}
\def\bt{\beta}
\def\mphi{m_{\phi}}
\def\hphi{\hat{\phi}}
\def\vphi{\langle \phi \rangle}
\def\etamunu{\eta^{\mu\nu}}
\def\dmul{\partial_{\mu}}
\def\dnul{\partial_{\nu}}
\def\bea{\begin{eqnarray}}
\def\eea{\end{eqnarray}}
\def\bfl{\begin{flushleft}}
\def\efl{\end{flushleft}}
\def\bc{\begin{center}}
\def\ec{\end{center}}
\def\bcr{\begin{center}}
\def\ecr{\end{center}}
\def\al{\alpha}
\def\bt{\beta}
\def\eps{\epsilon}
\def\lam{\lambda}
\def\gam{\gamma}
\def\s{\sigma}
\def\r{\rho}
\def\e{\eta}
\def\dl{\delta}
\def\non{\nonumber}
\def\nont{\noindent}
\def\la{\langle}
\def\ra{\rangle}
\def\nc{{N_c^{\rm eff}}}
\def\vp{\varepsilon}
\def\drho{\bar\rho}
\def\deta{\bar\eta}
\def\vma{{_{V-A}}}
\def\vpa{{_{V+A}}}
\def\J{{J/\psi}}
\def\ov{\overline}
\def\Lqcd{{\Lambda_{\rm QCD}}}
\def\btr{\bigtraingleup}


\def\issue(#1,#2,#3){{\bf #1}, #2 (#3)} 
\def\APP(#1,#2,#3){Acta Phys.\ Polon.\ \issue(#1,#2,#3)}
\def\ARNPS(#1,#2,#3){Ann.\ Rev.\ Nucl.\ Part.\ Sci.\ \issue(#1,#2,#3)}
\def\CPC(#1,#2,#3){Comp.\ Phys.\ Comm.\ \issue(#1,#2,#3)}
\def\CIP(#1,#2,#3){Comput.\ Phys.\ \issue(#1,#2,#3)}
\def\EPJC(#1,#2,#3){Eur.\ Phys.\ J.\ C\ \issue(#1,#2,#3)}
\def\EPJD(#1,#2,#3){Eur.\ Phys.\ J. Direct\ C\ \issue(#1,#2,#3)}
\def\IEEETNS(#1,#2,#3){IEEE Trans.\ Nucl.\ Sci.\ \issue(#1,#2,#3)}
\def\IJMP(#1,#2,#3){Int.\ J.\ Mod.\ Phys. \issue(#1,#2,#3)}
\def\JHEP(#1,#2,#3){J.\ High Energy Physics \issue(#1,#2,#3)}
\def\JPG(#1,#2,#3){J.\ Phys.\ G \issue(#1,#2,#3)}
\def\MPL(#1,#2,#3){Mod.\ Phys.\ Lett.\ \issue(#1,#2,#3)}
\def\NP(#1,#2,#3){Nucl.\ Phys.\ \issue(#1,#2,#3)}
\def\NIM(#1,#2,#3){Nucl.\ Instrum.\ Meth.\ \issue(#1,#2,#3)}
\def\PL(#1,#2,#3){Phys.\ Lett.\ \issue(#1,#2,#3)}
\def\PRD(#1,#2,#3){Phys.\ Rev.\ D \issue(#1,#2,#3)}
\def\PRL(#1,#2,#3){Phys.\ Rev.\ Lett.\ \issue(#1,#2,#3)}
\def\PTP(#1,#2,#3){Progs.\ Theo.\ Phys. \ \issue(#1,#2,#3)}
\def\RMP(#1,#2,#3){Rev.\ Mod.\ Phys.\ \issue(#1,#2,#3)}
\def\SJNP(#1,#2,#3){Sov.\ J. Nucl.\ Phys.\ \issue(#1,#2,#3)}
\def\ZPC(#1,#2,#3){Zeit.\ Phys.\ C \issue(#1,#2,#3)}

\def\pr{{\sl Phys. Rev.}~}
\def\prl{{\sl Phys. Rev. Lett.}~}
\def\pl{{\sl Phys. Lett.}~}
\def\np{{\sl Nucl. Phys.}~}
\def\zp{{\sl Z. Phys.}~}

\font\el=cmbx10 scaled \magstep2{\obeylines\hfill BITSGoa~-~2008/09/002}

\begin{center}

{\large \bf Implication of the HyperCP boson $X^0$ (214~MeV) in the flavour changing neutral current processes} 
\end{center}

\begin{center}
{\sl \large {Prasanta Kumar Das~\footnote{pdas@bits-goa.ac.in}}}
\end{center}

\begin{center}
Birla Institute of Technology and Science-Pilani, Goa campus\\
NH-17B, Zuarinagar, Goa 403726, India 
\end{center}


\centerline{\bf Abstract} {\small  We analyze the inclusive $b(c) \to s(u) \mu^+ \mu^-$ and the exclusive $B(D^+) \to K(\pi^+) \mu^+ \mu^-$ flavour changing neutral current decays in the light of HyperCP boson $X^0$ of mass $214$~MeV recently observed 
in the hyperon decay $\Sigma^+ \to p~ \mu^+ \mu^-$. Using the branching ratio data of the above inclusive and exclusive decays, we obtain constraints on $g_1~(h_1)$ and $g_2~(h_2)$, the scalar and pseudo-scalar coupling constants of the $b-s-X^0~(c-u-X^0)$ vertices. }

\bfl
{\it Keywords}: B-meson,~D-meson,~New Physics.\\
\vspace*{0.05in}
{\it PACS Nos.}: 14.40.Nd;~12.60.-i. 
\efl

\newpage
\section{Introduction}
The standard model(SM) despite of it's enormous experimental success, has 
some drawbacks arising from our little understanding of the quark and lepton 
flavour structures. The CKM matrix which tells about the mixing and the CP-violation in the quark sector, lacks any dynamical mechanism 
in it's origin. An urge for going beyond the SM by invoking non-standard new physics(NP) has become a driving force of the present-day phenomenological studies.  
Ideas like supersymmetry, technicolor, extra-dimension(s), little higgs model, unparticle physics, as a candidate of new physics, has drawn a lot of attention among the particle physics community. Experiments at colliders like Large Hadron Collider(LHC), upcoming International Linear Collider(ILC) will be the testing ground of all these novel ideas.  

 Recently the HyperCP collaboration found three events in the hyperon decay $\Sigma^+ \rightarrow p~ \mu^+ \mu^-$\cite{HyperCP}. In the di-muon invariant mass distribution plot (within the detector resolution) those were found to be localized at around $\sim 214$ MeV. The standard model alone cannot explain such distribution. To explain this they predict the existence of a new spin zero boson $X^0(214)$ which causes the flavour changing neutral current(FCNC) transition $s \to d X^0 $ followed by the 
HyperCP decay $X^0 \to \mu^+ \mu^-$. Usually in the standard model the FCNC processes are predominantly loop-mediated, so the tree-level FCNC process caused by this newly found $X^0$ boson spurs into a whole lot activity resulting several  interesting phenomenological studies \cite{DeshHe}. If such a boson $X^0$ indeed exists, one could also expect it to couple, besides $(s d)$ system, to $(c u)$, $(b d)$ and $(b s)$ systems. So far no study comprising the $X^0$ boson couplings to $(c u)$ and $(b s)$ are available. In the present work we explore these new couplings of the $X^0$ boson and discuss the experimental constraints on such couplings which  follows from the inclusive $b(c) \to s(u) \mu^+ \mu^-$ and exclusive $B(D^+) \to K(\pi^+) \mu^+ \mu^-$ decays, where the muon pair is produced from the decay of $X^0$ boson.

 We organize our paper in the following way. We analyze the inclusive $b\to s \mu^+ \mu^-$ and the exclusive $B \to K \mu^+ \mu^-$ decays within the SM and in the light of $X^0$ boson in section~\ref{sec:section2}. In section~\ref{sec:section3} we discuss in detail the inclusive decay $c \to u \mu^+ \mu^-$ and the exclusive decay $D^+ \to \pi^+ \mu^+ \mu^-$ in the SM and HyperCP scenarios.  
We discuss several  input parameters in section~\ref{sec:section4}. 
Section \ref{sec:section5} is devoted to the numerical analysis. We summarize and conclude in Section~\ref{sec:section6}.

\section{The inclusive and exclusive decays: $b \to s \mu^+ \mu^-$ and $B \to K \mu^+ \mu^-$ }
\label{sec:section2}

\subsection{The inclusive $b \to s \mu^+ \mu^-$ decay: SM and HyperCP analysis}
Within the SM the semi-leptonic inclusive decay $b \to s \mu^+ \mu^- $ comprising the FCNC transition is found to be dominated by the magnetic, electro-weak penguin and box operators. 

The effective weak Hamiltonian ${\cal H}_{\rm eff}$ for such $\Delta B = 1$ transition, can be written as \cite{bbl,Ali}
\bea\label{Eq:eff}
 {\cal H}_{\rm eff} = \frac{G_F}{\sqrt{2}} \left[ V_{jb} V_{js}^* \left(  c^{eff}_7 O_7 +c^{eff}_{9} O_{9} + c_{10} O_{10} \right)\right], ~~j=u, c, t
\eea
where the operators $O_{i}(~i=9,10)$ (semileptonic operators involving
electro-weak ($\gamma, Z$) penguin and box diagram)
and $O_7$ (magnetic penguin). The penguin and box diagrams

\begin{figure}[htb]
\begin{center}
\vspace*{1.2in}
      \relax\noindent\hskip -6.4in\relax{\includegraphics{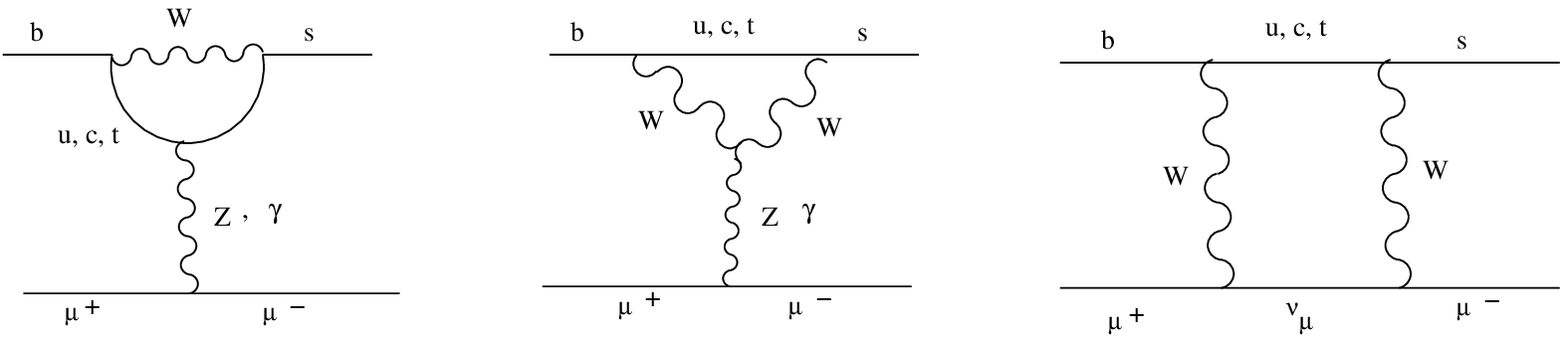}}
\end{center}
\end{figure}
\vspace*{-0.25in}
\noindent {Figure 1}.
{ \it Penguin and box diagrams contributing to 
$b(p_b) \to s(p_s) \mu^+ (p_1) \mu^- (p_2)$ process. }


\noindent which gives rise the above $b\to s \mu^+ \mu^-$ transition, are shown in Figure 1. The 
relevant operators are given by \cite{Ali}
\bea
O_9 &=& \frac{\alpha}{\pi} ({\overline{s}} \gamma_\mu P_L b) 
({\overline{\mu}}^- \gamma^\mu \mu^+),\nonumber \\
O_{10} &=& \frac{\alpha}{\pi} ({\overline{s}} \gamma_\mu P_L b) 
({\overline{\mu}}^- \gamma^\mu \gamma_5 \mu^+),\nonumber \\
O_{7} &=& \frac{\alpha}{\pi} ({\overline{s}} \sigma_{\mu\nu} q^\nu P_R b) 
\left[\frac{- 2 i m_b}{q^2}\right]({\overline{\mu}}^- \gamma^\mu \mu^+).
\eea
Here $\sigma_{\mu\nu} = \frac{i}{2} \left[\gamma_\mu,\gamma_\nu\right]$ and 
$P_{R,L} = \frac{1}{2} (1 \pm \gamma_5)$ are the chiral-projection operators.
$\alpha (= \frac{e^2}{4 \pi})$ is the QED fine structure constant.
The coefficients $c^{eff}_i~(i=7,9)$, known as the wilson coefficients, 
are evaluated at $\mu = m_b$. 
In above $q$ is the momentum transferred to the lepton pair and $m_b$ is 
the $b$-quark mass. We have neglected the term involving $m_s$(the $s$ quark mass) 
in operator $O_7$. 

 The squared amplitude ${\overline{|{\cal M}_{SM}|^2}}$ for the 
$b(p_b)\to s(p_s) \mu^+(p_1) \mu^-(p_2)$ decay within the SM  can be expressed as 
\bea \label{Eq:SMbsll}
{\overline{|M_{SM}|^2}} 
&=& \frac{1}{2} \sum_{spins} \left( |M_{7}|^2 + |M_{9}|^2 + |M_{10}|^2 + 
2 Re(M_9^* M_7) + 2 Re(M_{10}^* M_7) + 2 Re(M_9^*  M_{10}) \right) \nonumber \\
\eea
The direct and interference terms of the Eq.~(\ref{Eq:SMbsll}) are 
listed in Appendix A.1. 

Next we are to see the effect of the newly found spin zero HyperCP boson $X^0$ in the 
$b \to s \mu^+ \mu^-$ decay which can potentially be significant
at the tree level. The effective interaction describing such a FCNC decay as mediated by $X^0$ boson can be written as 
\bea
{ \cal L}_{d=4} = \left[{\overline s} (g_1 + i g_2 \gamma_5) b X^0 + h.c. \right]+ \left[ {\overline \mu} (l_1 + i l_2 \gamma_5) \mu X^0\right],
\eea
where $g_1$ and $g_2$ are the scalar and pseudo-scalar coupling constants 
of the $X^0$ boson with the $b$ and $s$ quarks and $l_1$ and $l_2$ are the scalar and pseudo-scalar coupling constants of the $X^0$ boson with the muon pairs. To find the HyperCP boson contribution, we work under the following assumption: 
we assume that the $X^0$ boson is produced as a {\it {real}} particle in the 
$b \to s X^0$ decay process and then decays to a muon pair $X^0 \to \mu^+ \mu^-$. The squared amplitude comprising the SM and NP contributions can be written as   
\bea \label{Eq:NPbsll}
{\overline{|M_{SM} + M_{NP}|^2}} ={\overline{|M_{SM}|^2}} +  {\overline{|M_{NP}|^2}}.
\eea
Several terms of Eq.~(\ref{Eq:NPbsll}) are given in  Appendix A.1 and A.2.  

 The decay width $\Gamma(b(p_b)\to s(p_s) \mu^+(p_1) \mu^-(p_2))_{SM}$ reads as
\bea \label{Eq:bsmmwid}
\Gamma(b\to s \mu^+ \mu^-)_{SM}=\frac{1}{512 \pi^3 m_b^3}\int_{(m_\mu + m_\mu)^2}^{(m_b-m_s)^2} \frac{d S_1}{S_1} \sqrt{\lambda_1 \lambda_2} \int_{-1}^{+1} dz~
\overline{|{ M}_{SM} (S_1,S_2(S_1,z))|^2},
\eea
where ${\overline{|M_{SM}|^2}}$ is given in Eq.~(\ref{Eq:SMbsll}). 
In Eq.(\ref{Eq:bsmmwid}) $z= cos\theta^\prime$, where 
$\theta^\prime$ is the angle in the center-of-mass frame of $\mu^+$ and 
$\mu^-$ and 
\bea
\lambda_1 &=& \lambda_1(m_b^2,m_s^2,S_1)=\sqrt{S_1^2 + m_b^4+m_s^4-2 S_1 m_b^2 - 2 S_1 m_s^2 - 2 m_b^2 m_s^2},\nonumber \\
\lambda_2 &=& \lambda(S_1,m_\mu^2,m_\mu^2)=\sqrt{S_1^2 -4 S_1 m_\mu^2}.
\eea 
In above $S_1 = q^2 = (p_b - p_s)^2$), $\lambda_i (i=1,2)$'s are the standard 
phase space $K{\ddot {a}}llen$ functions and 
$S_2 = A(S_1) + B(S_1)~cos\theta^\prime$, where 
\bea
A(S_1) &=& m_b^2 + m_\mu^2 - 2 m_b \gamma_X(S_1) \sqrt{m_\mu^2 + p^{\prime 2}(S_1)}, \nonumber \\
B(S_1) &=& -2 m_b  \gamma_X(S_1) \beta_X(S_1) p^{\prime}(S_1).
\eea
Here $p^{\prime}(S_1)=\sqrt{\frac{\lambda_2(S_1,m_\mu^2,m_\mu^2)}{4 S_1}}$, $\gamma_X=\frac{m_b^2 - m_s^2 + S_1}{2 m_b \sqrt{S_1}}$ and $\beta_X=\frac{\sqrt{\lambda_2(m_b^2,m_s^2,S_1)}}{m_b^2 - m_s^2 + S_1}$. Under the assumption of the real $X^0$ production, the NP contribution to the $\Gamma(b\to s \mu^+ \mu^-)_{NP}$ reads as 
\bea
\Gamma(b\to s \mu^+ \mu^-)_{NP} &=& \Gamma(b\to s X^0) \times BR[X^0 \to \mu^+ \mu^-] \nonumber \\
&=&\frac{1}{16 \pi m_b m_X^2} ({\overline{|M_{NP}|^2}}) \times BR[X^0 \to \mu^+ \mu^-],
\eea
where ${\overline{|M_{NP}|^2}}( = \frac{1}{2} \sum_{spins} |M_{NP}|^2)$ are 
given in Appendix A.2. 

\vspace*{0.25in}
\begin{figure}[htb]
\begin{center}
\vspace*{1.0in}
      \relax\noindent\hskip -3.4in\relax{\includegraphics{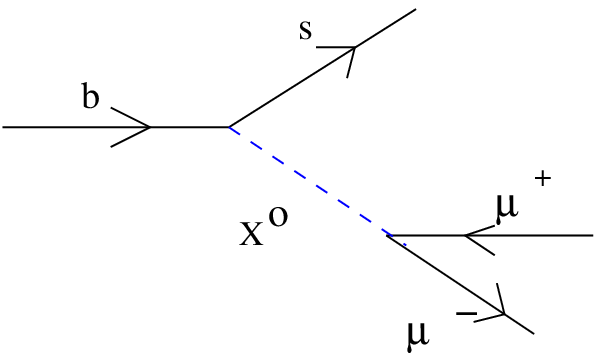}}
\end{center}
\end{figure}
\vspace*{-0.1in}
\noindent {Fig. 2}:
{ \it $X^0$ boson contribution to the inclusive decay $b \to s \mu^+(p_1) \mu^-(p_2)$. Momentum conservation reads $p_b = p_s +  p_1 + p_2$.}

\subsection{The exclusive decay $B \to K \mu^+ \mu^-$: SM and HyperCP analysis}
\label{sec:section3}
The exclusive semi-leptonic decay $B(p_B) \to K(p_K) \mu^+(p_1) \mu^-(p_2)$ (partonically it is the $b \to s \mu^+ \mu^- $ transition) is found to be dominated by the same set 
of operators (see Eq.~(\ref{Eq:eff})). The hadronic matrix elements necessary are \cite{Deshpande}
\bea  \label{Eq:mesonform1}
\langle K(p_K) |{\overline s} \gam_\mu  b| B(p_B)\rangle 
	  &=& \left[ (p_B + p_K)_\mu ~f_{BK}^+ (q^2) + q_\mu ~f_{BK}^- (q^2)\right],\\
\langle K(p_K) |{\overline s} \sigma_{\mu \nu} q^\nu b| B(p_B)\rangle 
             &=& \left[ q^2 ~(p_B + p_K)_\mu - (m_B^2 - m_K^2) ~q_\mu \right] 
f_{BK}^T (q^2),
\eea
where $q = p_B - p_K = p_1 + p_2$. Since $m_\mu \ll m_b$, the $q^\mu$ term in above equations gives negligible contribution. Working within the single pole with mass $\sim m_B$, the  $q^2$ dependence of the form factor can be written as 
\bea
f^+(q^2) = f^+(0)/(1 - q^2/m_B^2),~f^T(q^2) = f^T(0)/(1 - q^2/m_B^2).
\eea 
In the relativistic constituent quark model \cite{CQM}, we find 
\bea
f^+(0) \approx 0.34,~f^T(0) \approx f^+(0)/2m_b.
\eea
The full SM calculation gives the decay width
\bea \label{Eq:BKmmwidSM}
  \Gamma ( B \rightarrow K \mu^+ \mu^- ) = \frac{1}{512 \pi^3 m_B^3}\int_{(m_\mu + m_\mu)^2}^{(m_B-m_K)^2} \frac{d S'_1}{S'_1} \sqrt{\lambda_1 \lambda_2} \int_{-1}^{+1} dz~
\overline{|{ M'}_{SM} (S'_1,S'_2(S'_1,z))|^2} 
\eea
where $S'_1$, $S'_2$ are obtained from $S_1$ and $S_2$ by replacing $m_b$ and $m_s$ with $m_B$ and $m_K$ respectively. The squared amplitude ${\overline{|M'_{SM}|^2}}$  are given by  
\bea \label{Eq:SMBKmm}
{\overline{|M'_{SM}|^2}}
&=& \frac{1}{2} \sum_{spins} \left( |M'_{7}|^2 + |M'_{9}|^2 + |M'_{10}|^2 +
2 Re({M'_{9}}^* M'_{7}) + 2 Re({M'_{10}}^* M'_7) + 2 Re({M'_{9}}^*  M'_{10}) \right) \nonumber \\
\eea
The direct and interference terms of the Eq.~(\ref{Eq:SMBKmm}) are 
listed in Appendix B.1. 

 The Feynman diagram for the $X^0$ boson contribution to the 
$B \to K \mu^+ \mu^-$ decay is the same with the Figure 2 except that $b$ and $s$ quarks are to be replaced by $B$ and $K$ mesons. 
Considering the $X^o$ production as a real on-shell production, the 
squared-amplitude ${\overline{|M'_{NP}|^2}}$ can be written as 
\bea \label{Eq:NPBKmm}
{\overline{|M^{\prime}_{NP}|^2}} = \sum_{spins} |{M^\prime}_{NP}|^2,
\eea
where the direct term is given in Appendix B.2.  The $X^0$ boson contribution to the decay width $\Gamma(B \to K \mu^+ \mu^-)_{NP}$ is given by  
\bea \label{Eq:NPBKmmDW}
\Gamma(B \to K \mu^+ \mu^-)_{NP} &=& \Gamma(B \to K X^0) \times BR[X^0 \to \mu^+ \mu^-] \nonumber \\
&=&\frac{1}{16 \pi m_B m_X^2} (|M'_{NP}|^2) \times BR[X^0 \to \mu^+ \mu^-],
\eea
where $|M'_{NP}|^2$ is given in Appendix B.2.

\section{The inclusive and exclusive decays: $c \to u \mu^+ \mu^-$ and $D^+ \to \pi^+ \mu^+ \mu^-$ }
\label{sec:section3}
Although in the hadronic phenomena it is the downlike quark sector where the new physics effects(like tree level FCNC transition) are most likely to be seen, the uplike quark sector might equally be important.  
The HyperCP boson $X^0$, besides it's coupling with the $(b,s)$ system, can also couple with $(c,u)$ system and gives rise the FCNC trnsition in the up-like quark sector at the tree level. Although in general a charm meson decay is  dominated by the standard model long distance contribution, 
it is worthwhile to investigate the inclusive $c \to u \mu^+ \mu^-$ and the exclusive $D^+ \to \pi^+ \mu^+ \mu^-$ decays caused by the tree level FCNC transition $c \to u$ in the low di-muon invariant mass region.     

\subsection{The inclusive decay $c \to u \mu^+ \mu^-$: SM and HyperCP analysis}
For the inclusive $c \to u \mu^+ \mu^-$ decay it was found that the leading order rate is being suppressed by the QCD corrections within the Standard Model and a low dimuon $m_{\mu \mu}(= \sqrt{(p_1 + p_2})^2)$ have large impact on such decay. 

 The lagrangian describing the $c \to u \mu^+ \mu^-$ FCNC transition  can be written as 
 \bea\label{Eq:eff1}
 {\cal H'}_{\rm eff} = \frac{G_F}{\sqrt{2}} \left[ V_{cb}^* V_{ub} \left(  c^{'eff}_7 O'_7 +c^{'eff}_{9} O'_{9} + c'_{10} O'_{10} \right)\right],
\eea
where the operators are given by \cite{Fajfer}
\bea
O'_9 &=& \frac{\alpha}{\pi} ({\overline{u}} \gamma_\mu P_L c) 
({\overline{\mu}}^- \gamma^\mu \mu^+),\nonumber \\
O'_{10} &=& \frac{\alpha}{\pi} ({\overline{u}} \gamma_\mu P_L c) 
({\overline{\mu}}^- \gamma^\mu \gamma_5 \mu^+),\nonumber \\
O'_{7} &=& \frac{\alpha}{\pi} ({\overline{u}} \sigma_{\mu\nu} q^\nu P_R c) 
\left[\frac{- 2 i m_c}{q^2}\right]({\overline{\mu}}^- \gamma^\mu \mu^+)
\eea
and the effective wilson coefficients $c^{'eff}_7$, $c^{'eff}_9$ and 
$c'_{10}$ are evaluated at the scale $\mu = m_c$. The wilson coefficients required to evaluate the decay rate are given in the numerical analysis section. The decay rate is found to be  
\bea \label{Eq:cummwid}
\Gamma(c\to u \mu^+ \mu^-)_{SM}=\frac{1}{512 \pi^3 m_c^3}\int_{4 m_\mu^2)}^{(m_c-m_u)^2} \frac{d S_1}{S_1} \sqrt{\lambda_1 \lambda_2} \int_{-1}^{+1} dz~
\overline{|{ M^c}_{SM} (S_1,S_2(S_1,z))|^2}.
\eea
where the terms in amplitude square can be obtained from those obtained in section 2.1 by the following set of replacements:
$b \to c$, $s \to u$, $V_{tb}^* V_{ts} \to V_{cb}^* V_{ub}$, 
$m_b(m_s) \to m_c(m_u)$.  

 Next to find the HyperCP contribution to the inclusive 
 decay $c \to u \mu^+ \mu^-$. Although in the high di-muon invariant mass
such an effect is overshadowed by the long-distance contribution, in the low invariant mass region, one might single out the $X^0$ contribution. 
Treating $X^0$ production in $c \to u X^0$ decay as on-shell, the HyperCP contribution to the decay rate $\Gamma(c\to u \mu^+ \mu^-)_{NP}$
can be written as 
\bea
\Gamma(c\to u \mu^+ \mu^-)_{NP} &=& \Gamma(c\to u X^0) \times BR[X^0 \to \mu^+ \mu^-] \nonumber \\
&=&\frac{1}{16 \pi m_c m_X^2} ({\overline{|M^c_{NP}|^2}}) \times BR[X^0 \to \mu^+ \mu^-],
\eea
where ${\overline{|M^c_{NP}|^2}}( = \frac{1}{2} \sum_{spins} |M^c_{NP}|^2)$ are obtained from that give in Appendix A.2 simply by performing the following replacements: $p_b(p_s) \to p_c(p_u)$, $m_b(m_s) \to m_c(m_u)$ and $g_1(g_2) \to h_1(h_2)$ where $h_1$ and $h_2$ are
 the scalar and pseudoscalar couplings of the $c-u-X^0$ vertex).

\subsection{The exclusive decay $D^+ \to \pi^+ \mu^+ \mu^-$: SM and HyperCP analysis}
The exclusive $D^+ \to \pi^+ \mu^+ \mu^-$ decay rate is dominated by 
\cite{Fajfer} the long-distance resonant contributions at di-muon invariant mass $m_{\mu\mu}$ = $m_\rho(0.776),~m_\omega(0.782)$ and $m_\phi(1.02)$ (quantitities inside brackets indicates meson masses in GeV) and above this even the strongest NP contribution can not change the decay rate significantly. However in the low $m_{\mu \mu}$ region in the presence of a light scalar-pseudoscalar particle(the HyperCP boson $X^0(0.214)$) the situation might change and one can single out the HyperCP contribution to this exclusive semileptonic charm decay. The dynamics of this exclusive decay rate is governed by the same effective hamiltonian Eq.~(\ref{Eq:eff1}). Since the process is an exclusive one, the hadronic matrix elements required to be evaluated are \cite{Fajfer}
\bea  \label{Eq:mesonform2}
\langle \pi^+(p_\pi) |{\overline u} \gam_\mu  c| D^+(p_D)\rangle 
	  &=& \left[ (p_D + p_\pi)_\mu ~f_{D\pi}^+ (q^2) + q_\mu ~f_{D\pi}^- (q^2)\right],\\
\langle \pi^+(p_\pi) |{\overline u} \sigma_{\mu \nu} q^\nu c| D^+(p_D)\rangle 
             &=& \left[ q^2 ~(p_D + p_\pi)_\mu - (m_D^2 - m_\pi^2) ~q_\mu \right] f_{D\pi}^T (q^2),
\eea
where $q = p_D - p_\pi = p_1 + p_2$. Note that since $m_\mu \ll m_c$, the $q^\mu$ term in above two equations gives negligible contribution. As described before, we will be working within the single pole with mass $\sim m_D$ and the  $q^2$ dependence of the form factor can be written as 
\bea
f^+(q^2) = f^+(0)/(1 - q^2/m_D^2),~f^T(q^2) = f^T(0)/(1 - q^2/m_D^2).
\eea 
In the relativistic constituent quark model \cite{CQM}, we find 
\bea
f^+(0) \approx 0.73,~f^T(0) \approx f^+(0)/2m_c.
\eea
The decay width $\Gamma ( D^+ \rightarrow \pi^+ \mu^+ \mu^- )$ is obtained 
from Eq.~(\ref{Eq:BKmmwidSM}) simply by making the following replacements: $m_B(m_K) \to m_D(m_\pi)$ and $V_{tb} V_{ts}^* \to V_{cb}^* V_{ub}$ in the amplitude square.

The $X^0$ contribution to the decay rate $\Gamma ( D^+ \rightarrow \pi^+ \mu^+ \mu^- )$ is straightforward. In Eq.~(\ref{Eq:NPBKmmDW}) we replace $g_1(g_2)$ by $h_1(h_2)$ and the set of substitutions just mentioned above.  
\section{Input parameters} \label{sec:section4}
The decay rate depends on the 
CKM matrix elements, wilson coefficients, quark and lepton masses and the non-perturbative input e.g. form factors. 
\subsection{CKM matrix elements, quark masses, wilson coefficients and form factors}
We adopt the Wolfenstein parametrization with parameters $A,
\lam, \rho$ and $\eta$ of the CKM matrix as below
\bea
V_{CKM} =
\left( \begin{array}{ccc}
V_{ud} & V_{us}  & V_{ub} \\
V_{cd} & V_{cs}  & V_{cb} \\
V_{td} & V_{ts}  & V_{tb}\end{array} \right)
=\left(\begin{array}{ccc}
1 - \frac{1}{2}\lam^2 & \lam & A \lam^3 (\r - i \e) \\
-\lam & 1 - \frac{1}{2}\lam^2  & A \lam^2 \\
A \lam^3 (1 - \r- i \e) & - A \lam^2 & 1\end{array} \right).
\eea
 We set $A=0.815$ and $\lam(= \sin\theta_c)=0.2205$ in our analysis. 
Other relevant parameters are $ \r = \sqrt{{\overline \r}^2 + {\overline \eta}^2} 
~cos{\gamma}$ and $ \e = \sqrt{{\overline \r}^2 + {\overline \eta}^2}~ 
sin{\gamma}$, where $\sqrt{{\overline \r}^2 + {\overline \eta}^2}
= 0.3854$ and $\gamma \simeq 70^{o}$ \cite{PDG}.
For the quark masses we take their current value i.e.  $m_u=0.2,~m_d=0.2,~m_s=0.2,~m_c=1.5,~m_b=4.8,~m_t=175$ GeV \cite{Ali} and $m_\mu = 0.105$ GeV.


We have two set of wilson coefficients for studying the $B$ and $D$ decays. For the inclusive(exclusive) $b(B) \to s(K) \mu+ \mu-$ decay the following choice has been made: $c_7^{eff}=-0.313,~c_9^{eff}= 4.344$ and $c_{10}=-4.669$ (evaluated at $\mu = m_b$ in the NDR scheme) \cite{AliHiller}. On the other hand for the inclusive(exclusive) $c(D^+) \to u(\pi^+) \mu+ \mu-$ decay our choice is as follows: $c_7^{eff}=0.087,~c_9^{eff}= 
10^{-4}/(V^*_{cb} V_{ub})$ and $c_{10}=0$ (evaluated at $\mu = m_c$ in the NDR scheme)\cite{Fajfer}.

 For the form factors we will be working within the constituent quark model(CQM). 
In the present analysis we have used $f^+_{BK}(0) = 0.34$ \cite{Deshpande} and $f^+_{D\pi}(0) = 0.73$ (the central value as quoted in \cite{Fajfer}).  

\subsection{New boson $X^0$: it's coupling constants, mass and decay width}
A detailed analysis of the $X^0$ boson coupling to quarks and muon pairs is available 
in the literature \cite{DeshHe}. The observed $BR(K^+ \to \pi^+ \mu^+ \mu^-)$ which is about 
 ~$8.1 \times 10^{-8}$, imposes an upper bound on $|h'_1| < 7.4 \times 10^{-12}$ and a lower bound on $|h'_2| \ge 3.6 \times 10^{-10}$ \cite{DeshHe}. The present muon anomalous magnetic moment ($\delta a_\mu$) data imposes upper bound on 
$l_1$ and $l_2$ which are given by $|l_1|< 8.6 \times 10^{-4}$ and 
$|l_2|< 1.0 \times 10^{-3}$. The HyperCP event $\Sigma^+ \to p \mu^+ \mu^-$  suggests 
$m_X = 214$ (MeV). We set the $X^0$ boson decay width $\Gamma_X$ equal to zero. The 
$X^0$ boson which is produced as a real particle, decays to a muon pair only (i.e. $BR[X^0 \to \mu^+ \mu^-] = 1$), called Scenario I or can decay to a muon pair along with $X^0 \to \gamma \gamma, e^+ e^- $ (i.e. $BR[X^0 \to \mu^+ \mu^-] = 0.5$ say), called scenario II. 

\section{Numerical Analysis: ~Results and Discussions}\label{sec:section5}
In this section we obtain bounds on the scalar and pseudoscalar couplings by making the use of the inclusive and exclusive $B$ and $D$ decay data. First, we analyze the 
$B$ meson decay and then the decay of the $D$ meson.   

\subsection{SM and HyperCP analysis of $b \to s  \mu^+ \mu^-$ and $B \to K \mu^+ \mu^-$ decay}

In the standard model, the branching ratio for the inclusive 
$b\to s \mu^+ \mu^-$ decay is found to be $5.9 \times 10^{-6}$ 
and for the exclusive $B \to K \mu^+ \mu^-$ decay it is about $5.78 \times 10^{-7}$ \cite{AliHiller}. Since the branching ratio data of these inclusive and exclusive decays does not differ widely from the SM expectation, one would rather expect some stringent bounds on the couplings of the HyperCP boson $X^0$ with the SM fermions.  
\subsubsection{Analysis of the $b \to s  \mu^+ \mu^-$ decay }
The BELLE group found 
$BR(b\to s \mu^+ \mu^-) = 7.9 \pm 2.1 + 2.1(-1.5) \times 10^{-6}$  \cite{Belle} which is slightly larger than the SM value(see above). To see the effect of $X^0$ in $ b \to s X^0 (\to \mu^+ \mu^-)$ decay, we assume that $X^0$ boson is produced as a real on-shell particle which we already mentioned. The SM contribution which acts as a background, is considered only for the invariant mass interval $(m_X - 0.004)^2 \le S_1 \le (m_X + 0.40)^2$ in order to manifest the HyperCP boson effect prominent and is found to be  
\bea 
BR(b\to s \mu^+ \mu^-)_{SM} = 5.75 \times 10^{-7},
\eea
\noindent whic is one order smaller than the experimental result \cite{Belle}. This gives rise very tight constraints on the coupling constants $g_1$ and $g_2$ which 
we will see shortly. To obtain these constraints we develop the following strategy: 
we construct a quantity $\Delta_i$ defined as  $\Delta_i (=BR(b\to s \mu^+ \mu^-)_{expt} - BR(b\to s \mu^+ \mu^-)_{SM}$) (where $BR(b\to s \mu^+ \mu^-)_{SM}$ is evaluated with $(m_X - 0.004)^2 \le S_1 \le (m_X + 0.40)^2$). Note that $\Delta_i$ (by definition) is sensitive to the NP. We fit this quantity with the $X^0$ contribution and obtain constraints on the scalar $g_1$ and pseudoscalar $g_2$ coupling constants. 
In Figure 3a we have shown the contour plots in the $g_1 - g_2$ plane corresponding to  $\Delta_i$ ($i=0,1,2$) at the $0 \sigma$, $1\sigma$ and $2\sigma$ level by setting $BR[X^0] \to \mu^+ \mu^-= 1$ and similarly in Figure 3b corresponding to $BR[X^0] \to \mu^+ \mu^-= 0.5$. 
In Figures 3a and 3b, the region below the lowermost curve is allowed by 
$\Delta_0 (= BR(b\to s \mu^+ \mu^-)(C.V.) - BR(b\to s \mu^+ \mu^-)_{SM}) = 
1.998 \times 10^{-6}$, whereas the region lying below the middle curve is allowed by $\Delta_1 (= BR(b\to s \mu^+ \mu^-)(1\sigma) - BR(b\to s \mu^+ \mu^-)_{SM}) = 4.968 \times 10^{-6}$ and finally the region below the uppermost curve is allowed by $\Delta_2 (= BR(b\to s \mu^+ \mu^-)(2\sigma) - BR(b\to s \mu^+ \mu^-)_{SM}) = 7.938 \times 10^{-6}$. This gives rise the upper bounds on $g_1$ and $g_2$ which we will see next. From the intersection of the lowermost curve of Figure 3a with the $g_1$ and $g_2$ axes we finds $g_1 \le 2.4 \times 10^{-10}$ 
 and $g_2 \le 2.7 \times 10^{-10}$ and from the moddle one $g_1 \le 2.9 \times 10^{-10}$  and $g_2 \le 3.2 \times 10^{-10}$, respectively. From the uppermost curve it follows that $g_1 \le 3.3 \times 10^{-10}$  and $g_2 \le 3.6 \times 10^{-10}$. The upper bounds obtained on $g_1$ and $g_2$ from Figure 3b are as follows: the lowermost curve gives $g_1 \le 3.4 \times 10^{-10}$  and $g_2 \le 3.8 \times 10^{-10}$, the middle one $g_1 \le 4.1 \times 10^{-10}$  and $g_2 \le 4.5 \times 10^{-10}$ and finally the uppermost curve gives $g_1 \le 4.7 \times 10^{-10}$  and $g_2 \le 5.1 \times 10^{-10}$. 

\vspace*{-1.0in}
\newpage
\begin{figure}
\subfigure[]{
\label{PictureThreeLabel}
\hspace*{-0.7 in}
\begin{minipage}[b]{0.5\textwidth}
\centering
\includegraphics[width=\textwidth]{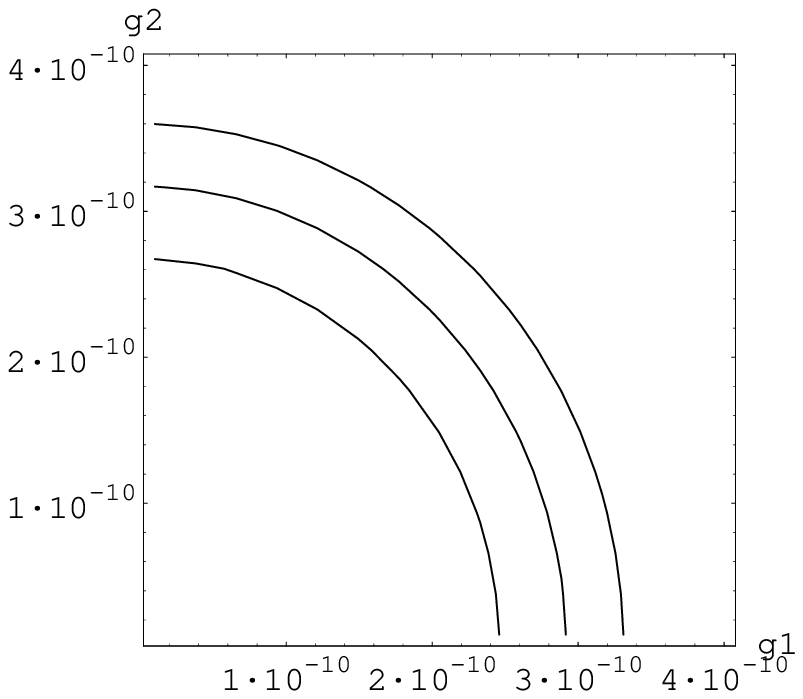}
\end{minipage}}
\subfigure[]{
\label{PictureFourLabel}
\hspace*{0.3in}
\begin{minipage}[b]{0.5\textwidth}
\centering
\includegraphics[width=\textwidth]{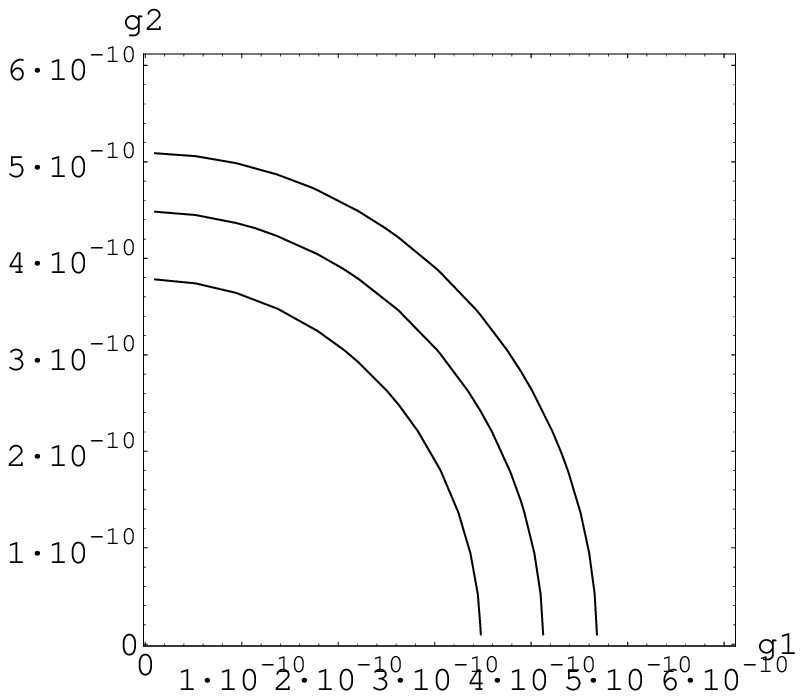}
\end{minipage}}
\end{figure}
\vspace*{-0.5in}
\noindent {\bf Figs. 3(a,b)}:
{{\it The $0\sigma$, $1\sigma$ and $2\sigma$  contour plots in the
scalar and pseudo-scalar couplings $g_1-g_2$ plane corresponding to $\Delta_i$($i=0,1,2$) (see the text for their definition) are shown in Figures  3a and 3b.  
In each Figure the uppermost, next to that(middle one) and the lowermost curve respectively stands for $\Delta_2 = 7.938 \times 10^{-6}$, $\Delta_1 = 4.968 \times 10^{-6}$ and  $\Delta_0 = 1.998 \times 10^{-6}$.Note that $BR[X^0 \to \mu^+ \mu^-]= 1.0$ and $0.5$ for Figures 3a and 3b, respectively.}}

\subsubsection{Analysis of the $B \to K  \mu^+ \mu^-$}
 The branching ratio $BR(B \to K \mu^+ \mu^-)$ within the SM is found to be $0.58 \times 10^{-6}$ and differs slightly from the data which is about  $(0.99^{+0.40~ +0.13}_{-0.32~ -0.14}) \times 10^{-6}$ \cite{BelleBK}.
In order to get an enhanced HyperCP effect, we consider the SM contribution with the invariant mass $S'_1$ as $(m_X - 0.004)^2 \le S'_1 \le (m_X + 0.40)^2$ and this gives 
\bea 
BR(B \to K \mu^+ \mu^-)_{SM} = 1.44 \times 10^{-8},
\eea
\noindent which is two order smaller than the experimental result \cite{BelleBK}. This gives rise very tight constraint on $g_1$ and $g_2$. We follow the strategy 
of the inclusive case. We define $\Delta_i (=BR(B\to K \mu^+ \mu^-)_{expt} - BR(B \to K \mu^+ \mu^-)_{SM}$). Note that as before $BR(B \to K \mu^+ \mu^-)_{SM}$ is evaluated with $(m_X - 0.004)^2 \le S'_1 \le (m_X + 0.40)^2$. Since $B$ and $K$ mesons are pseudo-scalar meson, one is able to constrain only the scalar coupling constant $g_1$. 
\vspace*{-1.0in}
\newpage
\begin{figure}
\subfigure[]{
\label{PictureThreeLabel}
\hspace*{-0.7 in}
\begin{minipage}[b]{0.5\textwidth}
\centering
\includegraphics[width=\textwidth]{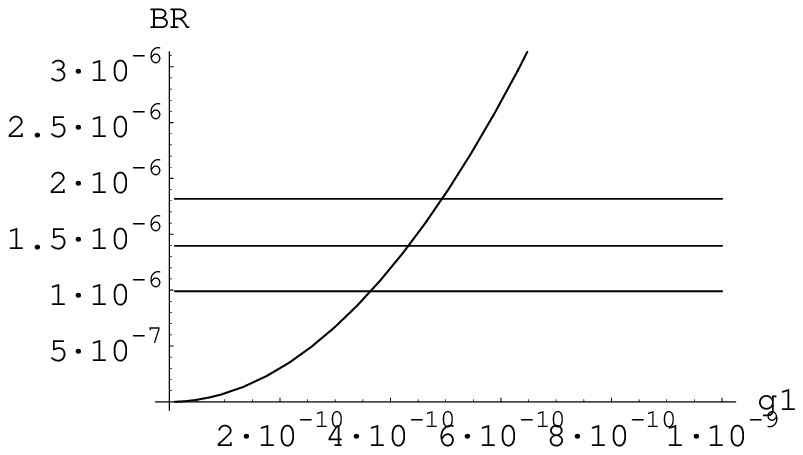}
\end{minipage}}
\subfigure[]{
\label{PictureFourLabel}
\hspace*{0.3in}
\begin{minipage}[b]{0.5\textwidth}
\centering
\includegraphics[width=\textwidth]{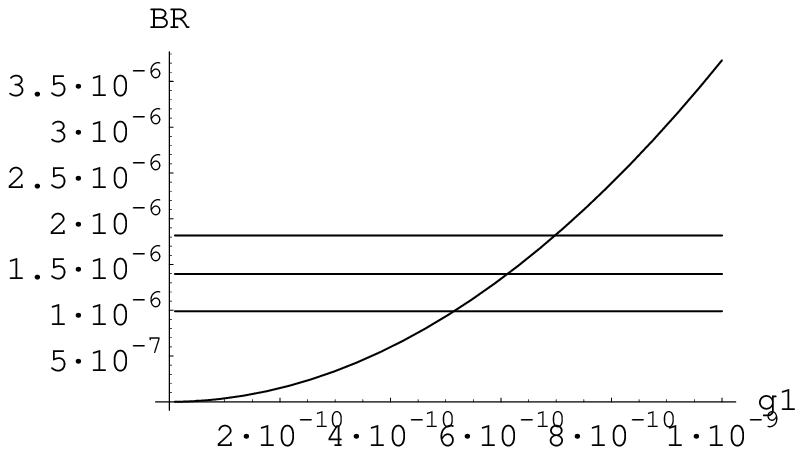}
\end{minipage}}
\end{figure}
\vspace*{-0.5in}
\noindent {\bf Figs. 3(c,d)}:
{{\it The $BR(B \to K \mu^+ \mu^-)_{NP}$ as a function of the scalar coupling constant $g_1$ is shown. The lower, middle and upper horizontal curves respectively stands for $\Delta_i$($i=0,1,2$) with $\Delta_2 = 1.8 \times 10^{-6}$, $\Delta_1 = 1.4 \times 10^{-6}$ and  $\Delta_0 = 0.99 \times 10^{-6}$. Note that $BR[X^0 \to \mu^+ \mu^-]= 1.0$ and $0.5$ for Figures 3c and 3d, respectively.}}

\noindent In Figures 3c and 3d we have plotted $BR(B \to K \mu^+ \mu^-)_{NP} (= BR(B \to K \mu^+ \mu^-)_{Expt} - BR(B \to K \mu^+ \mu^-)_{SM})$ as a function of $g_1$ corresponding $BR[X^0] \to \mu^+ \mu^-= 1$ and 
$BR[X^0] \to \mu^+ \mu^-= 0.5$, respectively. The lower, middle and upper horizontal  curves respectively stands for $\Delta_i$ where $i=0,1$ and $2$ with the following definitions: 
$\Delta_0 (= BR(B \to K \mu^+ \mu^-)(C.V.) - BR(B \to K \mu^+ \mu^-)_{SM})$,
$\Delta_1 (= BR(B \to K \mu^+ \mu^-)(1\sigma) - BR(B \to K \mu^+ \mu^-)_{SM}$,
and  
$\Delta_2 (= BR(B \to K \mu^+ \mu^-)(2\sigma) - BR(B \to K \mu^+ \mu^-)_{SM}$. The region below the horozontal curve and left side of the curve is allowed for $g_1$ and one obtain the upper bound 
$g_1 \le 3.8 \times 10^{-10}$ (from the $\Delta_0$ curve), $g_1 \le 4.33 \times 10^{-10}$ (from the $\Delta_1$ curve) and $g_1 \le  4.95 \times 10^{-10}$ (from the $\Delta_2$ curve), respectively. 

\subsection{SM and HyperCP analysis of the inclusive $c \to u  \mu^+ \mu^-$ and exclusive $D^+ \to \pi^+ \mu^+ \mu^-$ decay}
As we have seen in earlier subsection that within SM the long-distance resonance contribution controls the $D^+ \to \pi^+ \mu^+ \mu^-$ decay and the photon mediated penguin digram controls the $c \to u  \mu^+ \mu^-$ inclusive decay. To see the HyperCP  boson effect on the above exclusive decay where the SM resonance contribution can overshadow the HyperCP effect, we analyze this decay in the low invariant mass($S^c_1 = m_{\mu \mu}^2(=(p_1 + p_2)^2) < (0.7)^2 GeV^2$) region where the long-distance contribution gets suppressed. 

\vspace*{-1.0in}
\newpage
\begin{figure}
\subfigure[]{
\label{PictureThreeLabel}
\hspace*{-0.7 in}
\begin{minipage}[b]{0.5\textwidth}
\centering
\includegraphics[width=\textwidth]{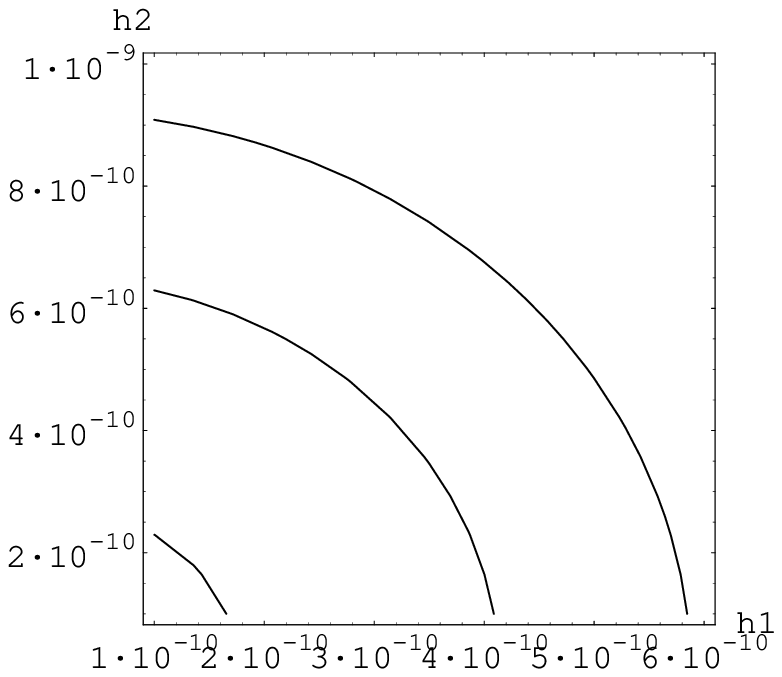}
\end{minipage}}
\subfigure[]{
\label{PictureFourLabel}
\hspace*{0.3in}
\begin{minipage}[b]{0.5\textwidth}
\centering
\includegraphics[width=\textwidth]{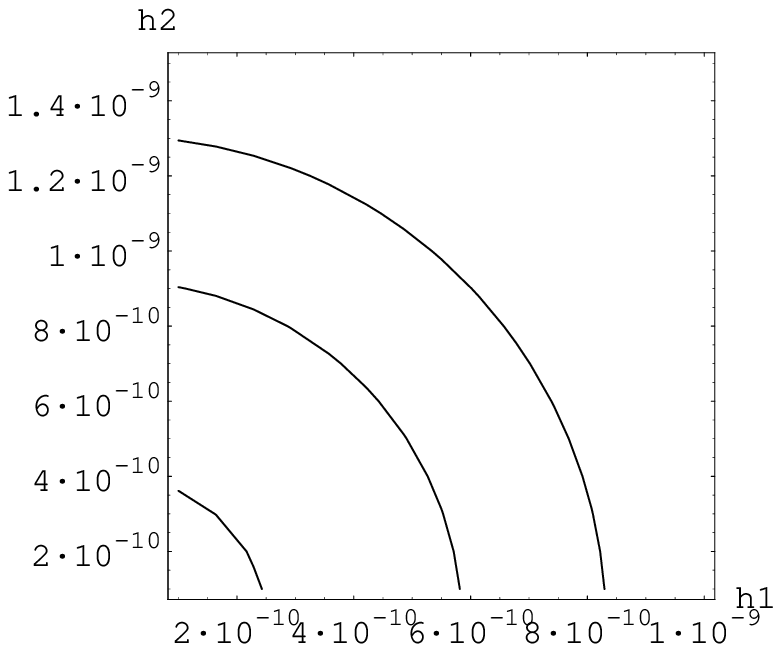}
\end{minipage}}
\end{figure}
\vspace*{-0.5in}
\noindent {\bf Figs. 3(e,f)}:
{{\it The contour plots in the scalar and pseudoscalar coupling constants $h_1 - h_2$  plane corresponding to different $R$ values are shown here. The lower, middle and upper horizontal curves respectively stands for $R = 10, 50$ and $100$. We set $BR[X^0 \to \mu^+ \mu^-]= 1.0$ and $0.5$, respectively in Figures 3e and 3f.}}
\subsubsection{Analysis of the $c \to u  \mu^+ \mu^-$}
For the SM  $c \to u  \mu^+ \mu^-$ decay rate we need the wilson coefficients $c^{eff}_i$($i=7,9,10$) which are listed in earlier section. Using those we find \cite{Fajfer} 
\bea 
BR(c\to u \mu^+ \mu^-)_{SM} = \frac{\Gamma^{SM}(c \to u \mu^+ \mu^-)}{\Gamma_{D^+}} = 1.15 \times 10^{-11}.
\eea
No data for the inclusive $c \to u  \mu^+ \mu^-$ decay rate so far is available. For our analysis what we do is as follows. First, we define 
the quantity
\bea
R = \frac{BR[c \to u  \mu^+ \mu^-]_{SM} + BR[c \to u  \mu^+ \mu^-]_{NP}}{BR[c \to u  \mu^+ \mu^-]_{SM}}.
\eea
Note that in the HyperCP effect in the $c \to u  \mu^+ \mu^-$ decay, we assume that the  $X^0$ boson is produced on-shell with 
$BR[X^0 \to \mu^+ \mu^-]= 1$ and $0.5$, respectively. Expecting that in future the  data will differs substantiably from the SM expectation, we obtain the contour plots in the $h_1$ and $h_2$ plane corresponding to $R=10,~50$ and $100$. These are shown in  Figures 3e and 3f. 
In Figures 3e and 3f we have set $BR[X^0 \to \mu^+ \mu^-]= 1.0$ and $0.5$, respectively. In each Figure,  the lower, middle and upper curves respectively corresponds to $R = 10$, $50$ and $100$. The region below the curve is allowed. From Figure 3e, corresponding to 
$R=10$ we find $h_1 \le 1.6 \times 10^{-10}$ and 
$h_2 \le 1.2 \times 10^{-10}$ and for $R=100$, we find 
$h_1 \le 4.0 \times 10^{-10}$ and 
$h_2 \le 6.76 \times 10^{-10}$. From Figure 3f, Corresponding to $R = 10$, we find $h_1 \le 2.0 \times 10^{-10}$ and 
$h_2 \le 2.4 \times 10^{-10}$ and for $R=100$, we find 
$h_1 \le 6.0 \times 10^{-10}$ and 
$h_2 \le 9.1 \times 10^{-10}$. 

\subsubsection{Analysis of the $D^+ \to \pi^+  \mu^+ \mu^-$}
The exclusive $D^+ \to \pi^+  \mu^+ \mu^-$ decay rate within the SM is found to be
largely  controlled by the long-distance resonance ($\rho, \omega$ and $\phi$ mesons) contribution. To study the $X^0$($m_X = 0.214$ GeV) impact on this exclusive decay we exclude those resonance contributions (background) by choosing $(2 m_\mu)^2 \le S^{c'}_1(=(m_{D^+} - m_{\pi^+})^2) \le (0.7)^2$ so that the HyperCP contribution does not get overshadowed by the SM one. We find the $BR(D^+ \to \pi^+  \mu^+ \mu^-)$ within the SM as
\bea 
BR(D^+ \to \pi^+  \mu^+ \mu^-)_{SM} = 7.85 \times 10^{-13},
\eea
which is much smaller than that obtained after including resonance contribution. For example, after the inclusion of the $\phi$ resonance one finds  
$BR(D^+ \to \phi \pi^+ \to \pi^+ \mu^+ \mu^-) = 
BR(D^+ \to \phi \pi^+) \times BR(\phi \to \mu^+ \mu^-) = 1.9 \times 10^{-6}$
\cite{Fajfer} which is comparable with the present experimental upper 
bound $8.8 \times 10^{-6}$. Anyway still there is a narrow window for the NP. Since both $D^+$ and $\pi^+$ are pseudo-scalar mesons, the upper bound only on the scalar HyperCP coupling $h_1$ is obtained. As before we define the quantity
\bea
R_1 = \frac{BR[D^+ \to \pi^+  \mu^+ \mu^-]_{SM} + BR[D^+ \to \pi^+  \mu^+ \mu^-]_{NP}}{BR[D^+ \to \pi^+  \mu^+ \mu^-]_{SM}}.
\eea
With the hope that in future the data will deviates from the SM result substantiably, we plot $R_1$ as a function of $h_1$. They are shown in Figures 3g and 3h. 
The lower, middle and upper horizontal lines in each Figures coresponds to  $R_1=10,~50$ and $100$. From Figure 3g we find $h_1 \le 0.615 \times 10^{-10},~1.425 \times 10^{-10}$ and $2.025 \times 10^{-10}$ corresponding
 to $R_1=10,50$ and $100$. On the other hand from Figure 3h we find $h_1 \le 0.87 \times 10^{-10},~2.015 \times 10^{-10}$ and $2.865 \times 10^{-10}$ corresponding to $R_1=10,50$ and $100$. No bound on the pseudo-scalar coupling $h_2$ is obtained from this exclusive decay.

\vspace*{-1.0in}
\newpage
\begin{figure}
\subfigure[]{
\label{PictureThreeLabel}
\hspace*{-0.7 in}
\begin{minipage}[b]{0.5\textwidth}
\centering
\includegraphics[width=\textwidth]{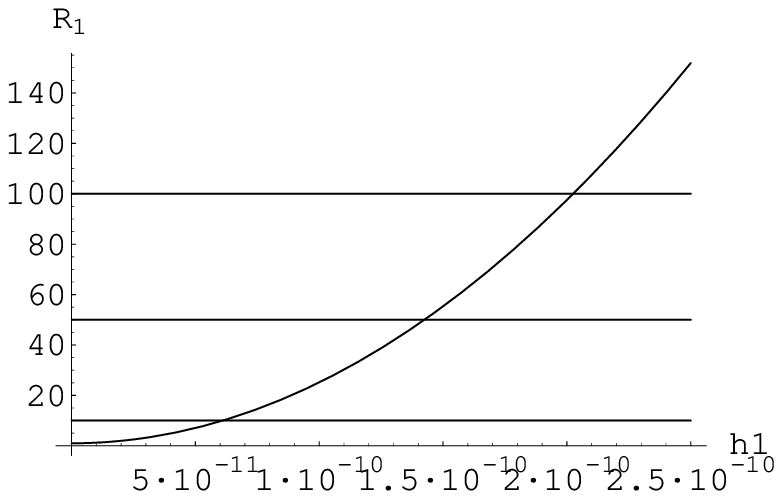}
\end{minipage}}
\subfigure[]{
\label{PictureFourLabel}
\hspace*{0.3in}
\begin{minipage}[b]{0.5\textwidth}
\centering
\includegraphics[width=\textwidth]{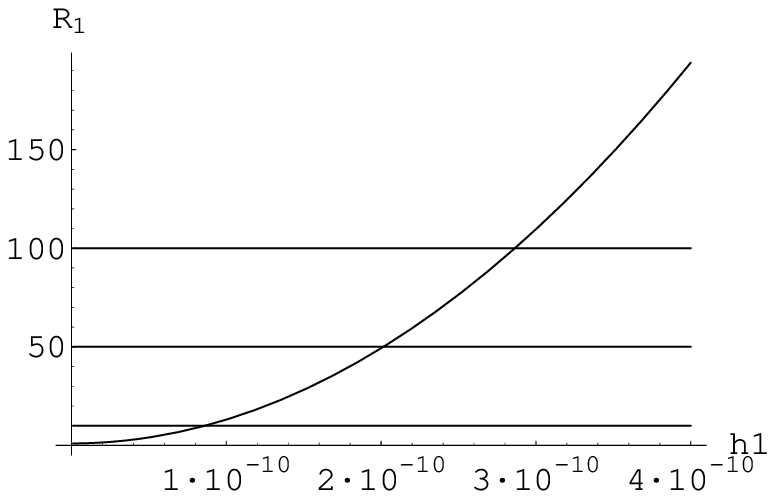}
\end{minipage}}
\end{figure}
\vspace*{-0.5in}
\noindent {\bf Figs. 3(g,h)}:
{{\it $R_1$ is plotted as a function of the scalar coupling constant $h_1$. The lower, middle and upper horizontal curves respectively stands for $R_1 = 10, 50$ and $100$. In Figures 3g and 3h we have set $BR[X^0 \to \mu^+ \mu^-]= 1.0$ and $0.5$, respectively.}}

\section{Summary and Conclusion} \label{sec:section6}
The HyperCP $X^0$ boson found in the $\Sigma^+ \to p \mu^+ \mu^-$ decay, besides it's coupling to the $(ds)$ system, can also couples to the $(bs)$ and $(cu)$ systems and this possibility is explored here. As a first case, we analyze the inclusive $b\to s \mu^+ \mu^-$ and the exclusive $B \to K  \mu^+ \mu^-$ decays in this HyperCP scenario and by using the experimental data, we have obtained the following upper bound on $g_1$ and $g_2$:  $g_1 \le 2.9 \times 10^{-10}$ and $g_2 \le 3.0 \times 10^{-10}$(from the inclusive data) and $3.8 \times 10^{-10}$ (from the exclusive data). From the incluse $c \to u \mu^+ \mu^-$ decay corresponding to $R=10$, we find $h_1 \le 2.0 \times 10^{-10}$ and  $h_2 \le 2.4 \times 10^{-10}$ and 
for $R=100$, we find  $h_1 \le 6.0 \times 10^{-10}$ and 
$h_2 \le 9.1 \times 10^{-10}$. Finally the exclusive  $D^+ \to \pi^+  \mu^+ \mu^-$ decay gives rise $h_1 \le 0.615 \times 10^{-10},~1.425 \times 10^{-10}$ and $2.025 \times 10^{-10}$ corresponding to $R_1=10,50$ and $100$, respectively assuming the $100 \%$ branching ratio of $X^0 \to \mu^- \mu^+$. 


\section{Acknowledgments}
The author would like to thank Prof. G.Rajasekaran and Prof. H.S.Mani of IMSc,Chennai for suggesting this problem and their valuable suggestions and comments from time to time throughout this work. He would also like to thank Prof. N.~G.~Deshpande of university of Oregon, USA for his useful comments on this work. 

\appendix
\section{ Calculation of the $b \to s \mu^+ \mu^-$ decay amplitude }
\subsection{SM amplitude for the $b(p_b) \to s(p_s) \mu^+(p_1) \mu^-(p_2)$ decay}

In this appendix, we calculate the square of the amplitude of Eq.~(\ref{Eq:SMbsll}).
We define $p_b$, $p_s$, $p_1$ and $p_2$ to be the momenta of the b-quark, 
s-quark, $\mu^+$ and $\mu^-$, respectively with $q = p_b -p_s = p_1 + p_2$.
The individual amplitude-square elements are given by 
\bea
|M_7|^2=\left(\frac{\alpha G_{F}}{\pi\sqrt{2}}\right)^2 |V_{tb}|^{2} 
|V_{ts}|^{2}|C_{7}^{eff}|^{2}(\frac{2m_{b}}{q^{2}})^{2}
Tr\left[(\slash{p_{b}}+m_{b})\sigma_{\rho\alpha}q^{\alpha}P_{L}
(\slash{p_{s}}+m_{s})\sigma_{\mu\nu}q^{\nu}P_{R}\right] \times
\nonumber \\
Tr\left[(\slash{p_{1}}-m_{\mu})\gamma^{\rho}(\slash{p_{2}}+m_{\mu})\gamma^{\mu}\right], \nonumber \\
|M_9|^2=\left(\frac{\alpha G_{F}}{\pi\sqrt{2}}\right)^{2} |V_{tb}|^{2} 
|V_{ts}|^{2}|C_{9}^{eff}|^{2}Tr\left[(\slash{p_{b}}+m_{b})\gamma_{\nu}P_{L}(\slash{p_{s}}+m_{s})\gamma_{\mu}P_{L}\right]  \times
\nonumber \\
Tr\left[(\slash{p_{1}}-m_{\mu})\gamma^{\nu}(\slash{p_{2}}+m_{\mu})\gamma^{\mu}\right],\nonumber \\
|M_{10}|^2=\left(\frac{\alpha G_{F}}{\pi\sqrt{2}}\right)^{2} |V_{tb}|^{2} 
|V_{ts}|^{2}|C_{10}|^{2}Tr\left[(\slash{p_{b}}+m_{b})\gamma_{\nu}P_{L}(\slash{p_{s}}+m_{s})\gamma_{\mu}P_{L}\right]  
\times
\nonumber \\
Tr\left[(\slash{p_{1}}-m_{\mu})\gamma^{\nu}\gamma^{5}(\slash{p_{2}}+m_{\mu})\gamma^{\mu}\gamma^{5}\right], \nonumber \\
2Re(M_9^* M_7)=2\left(\frac{\alpha G_F}{\pi\sqrt{2}}\right)^{2} |V_{tb}|^{2} 
|V_{ts}|^{2} Re((C_{9}^{*eff}C_{7}^{eff})(\frac{-2im_{b}}{q^{2}})Tr[(\slash{p_{s}}+m_{s})\sigma_{\mu\rho}q^{\rho}P_{R}(\slash{p_{b}}+m_{b})\gamma_{\nu}P_{L}
\nonumber \\
\times Tr[(\slash{p_2}+m_\mu) \gam^\mu (\slash{p_1}-m_\mu) \gam^\nu] )
,\nonumber \\ 
2Re( M_9^* M_{10})=2\left(\frac{\alpha G_{F}}{\pi\sqrt{2}}\right)^{2} |V_{tb}|^{2} |V_{ts}|^{2} Re((C_{9}^{*eff}C_{10})Tr\left[(\slash{p_{s}}+m_{s})\gamma_{\mu}P_{L}(\slash{p_{b}}+m_{b})\gamma_{\nu}P_{L}\right]
 \times
\nonumber \\
Tr\left[(\slash{p_{2}}+m_{\mu})\gamma^{\mu}\gamma^{5}(\slash{p_{1}}-m_{\mu})\gamma^{\nu}\right]), \nonumber \\
2Re(M_{10}^* M_7)=2\left(\frac{\alpha G_{F}}{\pi\sqrt{2}}\right)^{2} |V_{tb}|^{2} |V_{ts}|^{2}Re((C_{10}^{*}C_{7}^{eff})(\frac{-2im_{b}}{q^{2}})Tr\left[(\slash{p_{s}}+m_{s})\sigma_{\mu\rho}q^{\rho}P_{R}(\slash{p_{b}}+m_{b})\gamma_{\nu}P_{L}\right]
 \times
\nonumber \\
Tr\left[(\slash{p_{2}}+m_{\mu})\gamma^{\mu}(\slash{p_{1}}-m_{\mu})\gamma^{\nu}\gamma^{5}\right]). \nonumber 
\eea
\subsection{NP amplitude for the $b(p_b) \to s(p_s) X^0(p_X)$ decay}

The amplitude-square element of  Eq.~(\ref{Eq:NPbsll}) is given by 
\bea
|M_{NP}|^2=
4 \left((g_1^2 - g_2^2) m_b m_s + (g_1^2 + g_2^2) p_b.p_s\right). 
\nonumber \\
\eea
\section{ $B(p_B) \to  K(p_K) \mu^+(p_1) \mu^-(p_2)$ decay amplitudes }
\subsection{Standard Model terms}
In this appendix, we calculate the square of the SM amplitude of 
$B (p_B) \rightarrow K (p_K) \mu^+(p_1) \mu^-(p_2)$ of 
Eq.~(\ref{Eq:SMBKmm}).
Here $q = p_B - p_K = p_1 + p_2$. The individual terms are given by\\
\bea
|M'_7|^2=\left(\frac{\alpha G_{F}}{\pi\sqrt{2}}\right)^2 |V_{tb}|^{2} 
|V_{ts}|^{2}|C_{7}^{eff}|^{2}(\frac{2m_{b}}{q^{2}})^{2}
Tr\left[(\slash{p_{2}}+m_{\mu})\gamma^{\mu}(\slash{p_{1}}-m_{\mu})\gamma^{\nu}\right] \times
\nonumber \\
\frac{1}{4} q^4 (p_B + p_K)_\mu (p_B + p_K)_\nu |f^T_{BK}(q^2)|^2, \nonumber \\
|M'_9|^2=\left(\frac{\alpha G_{F}}{\pi\sqrt{2}}\right)^{2} |V_{tb}|^{2} 
|V_{ts}|^{2}|C_{9}^{eff}|^{2} Tr\left[(\slash{p_{1}}-m_{\mu})\gam^{\nu}(\slash{p_{2}}+m_{\mu})\gam^{\mu}\right]  \times
\nonumber \\
\frac{1}{4} (p_B + p_K)_\mu (p_B + p_K)_\nu |f^+_{BK}(q^2)|^2,\nonumber \\
|M'_{10}|^2=\left(\frac{\alpha G_{F}}{\pi\sqrt{2}}\right)^{2} |V_{tb}|^{2} 
|V_{ts}|^{2}|C_{10}|^{2}Tr\left[(\slash{p_{1}}-m_{\mu})\gam^{\nu} \gam^5(\slash{p_{2}}+m_{\mu})\gam^{\mu} \gam^5\right]  \times
\nonumber \\
\frac{1}{4} (p_B + p_K)_\mu (p_B + p_K)_\nu |f^+_{BK}(q^2)|^2,\nonumber \\
2 Re({M'_9}^* M'_7)=2\left(\frac{\alpha G_F}{\pi\sqrt{2}}\right)^{2} |V_{tb}|^{2} 
|V_{ts}|^{2} Re((C_{9}^{*eff}C_{7}^{eff})(\frac{-2im_{b}}{q^{2}})~ 
Tr[(\slash{p_2}+m_\mu) \gam^\mu (\slash{p_1}-m_\mu) \gam^\nu] ) \times \nonumber \\
\frac{1}{4} q^2 (p_B + p_K)_\mu (p_B + p_K)_\nu f^+_{BK}(q^2) 
f^T_{BK}(q^2) ,\nonumber \\ 
2Re( M'_9 {M'_{10}}^*)=2\left(\frac{\alpha G_{F}}{\pi\sqrt{2}}\right)^{2} |V_{tb}|^{2} |V_{ts}|^{2} Re((C_{9}^{eff}C_{10}^*)~ Tr[(\slash{p_1}-m_\mu) \gam^\nu (\slash{p_2}+m_\mu) \gam^\mu \gam^5] ) \times \nonumber \\
\frac{1}{4} (p_B + p_K)_\mu (p_B + p_K)_\nu |f^+_{BK}(q^2)|^2,\nonumber \\ 
2Re({M'_{10}}^* M'_7)=2\left(\frac{\alpha G_{F}}{\pi\sqrt{2}}\right)^{2} |V_{tb}|^{2} |V_{ts}|^{2}Re((C_{10}^{*}C_{7}^{eff})(\frac{-2im_{b}}{q^{2}})~Tr\left[(\slash{p_{1}}-m_{\mu})\gam^{\nu}(\slash{p_{2}}+m_{\mu})\gam^{\mu}\gam^{5}\right]) \times
\nonumber \\
\frac{1}{4} q^2 (p_B + p_K)_\mu (p_B + p_K)_\nu f^+_{BK}(q^2) 
f^T_{BK}(q^2). \nonumber 
\eea

\subsection{$X^0$ boson contribution}

The $X^0$ boson contribution (Eq.~(\ref{Eq:NPBKmm})) can be written as 
\bea
| M'_{NP}|^2= \frac{g_1^2}{m_b} (m_B^2 - m_K^2)^2 ~|f^+_{BK}|^2.
\eea


\end{document}